\documentstyle[12pt]{article}
\def\be{\begin{equation}}  \def\ee{\end{equation}}   
\def\bea{\begin{eqnarray}} \def\eea{\end{eqnarray}}   

\def\CC{\kern 0.27em \vrule height1.45ex width0.03em depth0em \kern-0.30em%
\rm C}
\newcommand{\ignore}[1]{}

\def\cC{{\cal C}}

\def\nn{\nonumber}

\begin{document}

\hoffset = -1truecm
\voffset = -2truecm


\bibliographystyle{plain}

\title{
A N-extended version of superalgebras in $D=3,\; 9 \;{\rm mod}\; 8$ 
\thanks{This work was partially supported by CONICET, Argentina} 
}


\author{ Adri\'an R. Lugo
\thanks{Electronic address: lugo@dartagnan.fisica.unlp.edu.ar}\\
\normalsize Departamento de F\'\i  sica, Facultad de Ciencias Exactas\\
\normalsize Universidad Nacional de La Plata\\C.C. 67, (1900) La Plata,
Argentina
}

\date{September 1997}

\maketitle
\begin{abstract}
A set of generalized superalgebras containing arbitrary tensor
p-form operators is considered in dimensions $D=2n+1$ for 
$n=1,4 {\;\rm mod}\; 4$ and the general conditions for its existence
expressed in the form of generalized Jacobi identities is established. 
These are then solved in a univoque way and some lowest dimensional cases
$\; D= 3, 9, 11\;$of possible interest are made explicit. 

\end{abstract}
\bigskip

\section*{Introduction}

Supersymmetry (SUSY) was a deeply studied subject during the last two
decades since the no-go theorem of Coleman and Mandula about all the
possible symmetries of the S-matrix in $D=4$ was by-passed by the
discovery of Haag, Lopuzsanski and Sohnius (HLS) on possible fermionic
extensions of the corresponding symmetry algebra \cite{susy}.
While I do not know the relevance of the mentioned theorem in
dimensions greater than four and outside the domain of local quantum
field theory (and the present paper do not intend to address this question) 
it is certainly true that SUSY in higher dimensions becamed more and more
relevant with the construction of the various supergravity theories in the
seventies \cite{sugra}, with the formulation of superstrings theories in
the eighties \cite{sustrings} and with the more recently conjectured
existence of a hypothetic M-theory in $D=11$\cite{mtschwarz}.
And in relation with this last one and  possible compactifications of
it that give rise to SUSY theories in lesser dimensions, it naturelly 
appear ``charged" extensions of the superalgebras considered before that
were not allowed by the HLS theorem. 
These charges in fact present non trivial Lorentz transformation properties 
and were found as topological charges associated to solutions of
$p$-branes in supergravity theories \cite{stelle}, being given
essentially by generalizations of the familiar string winding number 
\be
Z^{M_1 \dots M_p } \sim \int_{\Sigma_p} \; dX^{M_1} \wedge \dots 
\wedge d X^{M_p} \label{cargas}
\ee 
where $\Sigma_p$ is the spatial volume of the $p$-brane world volume.
A nonzero value of $Z$ is obtained if at some fixed proper time the 
$p$-brane defines a non trivial $p$-cycle in space-time. 

In references \cite{holpro} , \cite{azca}, \cite{bergsez}, \cite{sez1} ,
some possible generalizations of superalgebras containing p-form
generators of the kind given in (\ref{cargas}) and additional spinorial ones 
(generalized $p$-forms in superspace) were considered, mainly focused in
the minimal $N=1$ case and in eleven dimensions. 
In this letter a $N$-extended version of superalgebras in 
$D= 3,\; 9,\;\rm{mod}\; 8$ containing only tensor generators of the
type (\ref{cargas}) is presented.

Two appendices are included which contain the definitions and formulae 
used throughout the paper.

\section*{The extended superalgebras}

I start by considering the existence of a $\rm Spin(1,2n)$ algebra with
generators $\{ X_{MN} = - X_{NM} , M = 0,1,\dots , 2n \}$ satisfying the 
standard commutation relations
\be
\left[ X_{MN}, X_{PQ} \right] = \eta_{MQ}\; X_{NP} + \eta_{NP}\; X_{MQ} 
                              - \eta_{MP}\; X_{NQ} - \eta_{NQ}\; X_{MP} 
\label{loralg}
\ee 
For $n=1, 4 \; \rm{mod}\; 4$ it is possible to add to these bosonic
generators $N$ anticommuting Majorana supercharges 
$\{ Q_I^\Lambda ; I=1,\dots,N , \Lambda = 1,\dots, 2^n \}$,
\be
\left[ Q_I^\Lambda , X_{MN}\right] = {(S_{MN})}^\Lambda{}_\Omega\;
Q_I^\Omega\label{qxq}
\ee
where the $S_{MN}$'s are the spinor representation of the $X_{MN}$'s and 
are defined in Appendix A.
The completeness of the $\Gamma^{(\mu_p )}$'s allows to write the
general anticommutation relations
\be
\left\{ Q_I^\Lambda , Q_J^\Omega \right\} = \sum_{p=0}^n\;
\frac{(-)^{\frac{p}{2}(p-1)}}{p!}
\;( \Gamma^{(\mu_p )} \cC^{-1} )^{\Lambda\Omega}\; 
Z_{IJ(\mu_p )}^{(p)}\label{qq}
\ee
where due to (\ref{qxq}) and the XQQ superJacobi identity 
the $\{Z_{IJ(\mu_p )}^{(p)} \}$ is a set of $p$-form valued commuting 
operators, e. g. for any  ${\rm Spin}(1, 2n)$ element 
$\; U(\omega) = e^{{1\over 2}\omega^{MN} X_{MN}}$,
\bea
U(\omega)^{-1}\; Z_{IJ(\mu_p )}^{(p)}\; U(\omega) &=&  
(V(\omega)^{-1})^{(\nu_p )}{}_{(\mu_p )} \; Z_{IJ(\nu_p )}^{(p)}\nn\\
V(\omega )^{(\mu_p)}{}_{(\rho_p)}\; (V(\omega )^{-1} )^{(\rho_p)}{}_{(\nu_p)} 
&=& \delta^{(\mu_p)}{}_{(\nu_p)}  
\eea
where the $V$ and $\delta$ are introduced in (\ref{deltadef}) and
(\ref{vdef}).
By consistency they must satisfy the symmetry property 
\be
Z_{IJ(\mu_p )}^{(p)} = \delta_p\; Z_{JI(\mu_p )}^{(p)}\label{simZ}
\ee
where $\delta_p$ is the phase introduced in (\ref{gamaprop1}).
On the other hand, admitting grading of the algebra the XZQ superJacobi
identity 
and equation (\ref{gamalor}) dictate the commutation relation
\be
\left[ Q_K^\Lambda , Z_{IJ(\mu_p )}^{(p)}\right] =
\omega^{(p)}_{IJK}{}^L \; ({\Gamma_{(\mu_p )}})^\Lambda{}_\Omega\;Q_L^\Omega
\label{qz}
\ee
where the coefficients $\omega^{(p)}_{IJK}{}^L$ (written some times 
as matrix elements) must satisfy 
\be
\omega^{(p)}_{IJK}{}^L \equiv (\omega^{(p)}_{IJ})_K{}^L 
= \delta_p \; (\omega^{(p)}_{JI})_K{}^L \label{simw}
\ee
It will be shown they determine the whole algebra.

\section*{The super-Jacobi identities}

The consistency with the generalized Jacobi identities impose as usual
strong constraints on the the possible superalgebras. 
Those identities involving Lorentz generators were used above to write the
algebra, here the remaining ones will be imposed.
  
To start with the $ZQQ$ identity 
is written as
\bea
0 &=& \sum_{r=0}^n \frac{(-)^{\frac{r}{2}(r-1)}}{r!}
\left( \omega^{(p)}_{IJK}{}^E \; \delta_L^F 
\Gamma_{(\mu_p )} \;\Gamma^{(\rho_r )}
+ \omega^{(p)}_{IJL}{}^E \; \delta^F_K 
\left( \Gamma_{(\mu_p )}\Gamma^{(\rho_r )}\;\cC^{-1} \right)^t \cC 
\right) Z_{EF(\rho_r )}^{(r)}\nn\\
&+&\sum_{q=0}^n
\frac{(-)^{\frac{q}{2}(q-1)}}{q!}\left[ Z_{IJ(\mu_p )}^{(p)},
Z_{KL(\nu_q)}^{(q)}\right]\;\Gamma^{(\nu_q)}  
\label{j1}
\eea
On the other hand the $ZZQ$ identity 
becomes
\bea
\left[ Q_M^\Lambda, \left[ Z_{IJ(\mu_p )}^{(p)},
Z_{KL(\nu_q)}^{(q)}\right]\right] &=&   
\biggl( (\omega^{(p)}_{IJ}\; \omega^{(q)}_{KL})_M{}^E \;
( \Gamma_{(\mu_p )} \; \Gamma_{(\nu_q )} )^\Lambda{}_\Delta\\
&-&(\omega^{(q)}_{KL} \;\omega^{(p)}_{IJ})_M{}^E\;
( \Gamma_{(\nu_q)} \; \Gamma_{(\mu_p )} )^\Lambda{}_\Delta \biggr)\;
Q_E^\Delta \label{j2}   
\eea
Finally the $QQQ$ identity 
yields
\begin{eqnarray}
0 &=& \sum_{q=0}^n  \frac{(-)^{\frac{q}{2}(q-1)}}{q!}\;
\biggl(\omega^{(q)}_{IJK}{}^L\;
(\; \Gamma^{(\nu_q)} \cC^{-1} )^{\Lambda\Omega}\;
(\Gamma_{(\nu_q)} \cC^{-1} )^{\Delta\Upsilon} \nn\\
&+&\omega^{(q)}_{KIJ}{}^L\;
(\Gamma^{(\nu_q)} \cC^{-1} )^{\Delta\Lambda}\;
(\Gamma_{(\nu_q)} \cC^{-1} )^{\Omega\Upsilon} 
+ \omega^{(q)}_{JKI}{}^L\;(\Gamma^{(\nu_q)} \cC^{-1})^{\Omega\Delta}
(\Gamma_{(\nu_q)} \cC^{-1} )^{\Lambda\Upsilon} \biggr)  \label{j3}
\end{eqnarray}
The use of completeness of the $\Gamma^{(\mu_p)}$'s and some
matrix relations quoted in the appendices give simplifyied versions of
these conditions. 
In first term let us consider (\ref{j1}); then equations
(\ref{2gamabis}) and (\ref{simcC}) {\it fix} the $Z$-algebra to be
\bea
\left[ Z_{IJ(\mu_p )}^{(p)}, Z_{KL(\nu_q)}^{(q)}\right] &=& 
\sum_{r=0}^n \; \frac{1}{r!}\;
{\cal Q}_{IJKL}^{EF}(p;q,r)\; \cC^{(\rho_r)}_{(\mu_p)(\nu_q)} 
Z_{EF(\rho_r )}^{(r)} \nn\\
{\cal Q}_{IJKL}^{EF}(p;q,r) &=& -{1\over 2}\; \left(
\omega^{(p)}_{IJK}{}^E \;\delta_L^F + 
\delta_q\;\omega^{(p)}_{IJL}{}^E \;\delta_K^F \right) +
\delta_r\;\left( E\leftrightarrow F\right)
\label{zzz}
\eea
Not only this, consistency with the antisymmetry of the commutator gives
the constraint
\be
{\cal Q}_{IJKL}^{EF}(p;q,r) + \sigma_{pqr}\; {\cal Q}_{KLIJ}^{EF}(q;p,r) 
= 0\label{j12}
\ee

In second term the same equations allow to rewrite
(\ref{j2}) as a constraint on the $\omega^{(p)}$'s
\be
\omega^{(p)}_{IJ}\; \omega^{(q)}_{KL} - \sigma_{pqr}\;
\omega^{(q)}_{KL} \;\omega^{(p)}_{IJ} =
{\cal Q}_{KLIJ}^{EF}(q;p,r) \; \omega^{(r)}_{EF} \label{j22}
\ee

Finally by using the identity (\ref{gamaprop2}) equation (\ref{j3}) 
becomes equivalent to
\footnote{
For $n=1$ equation (\ref{gamaprop2}) does not work, however (\ref{j32})
remains valid.
}
\be
2^n\;\omega^{(0)}_{KJI}{}^L + \sum_{p=0}^n  
{2n+1\choose p}\; \left( \omega^{(p)}_{IJK}{}^L + 
\eta\;\omega^{(p)}_{IKJ}{}^L \right)= 0 \label{j32}
\ee

In the next section the set of equations (\ref{j12}), (\ref{j22}), 
(\ref{j32}) will be solved.

\section*{The general solution}
\bigskip

Let us start by considering the right hand side of (\ref{j22}) at fixed 
$p$ and $q$ for two different $r, r'\;$ such that 
$\sigma_{pqr}= \sigma_{pqr'}$ (and then $\delta_r = \delta_{r'}$); 
it follows that
\be
{\cal Q}_{IJKL}^{EF}(p;q,r) = {\cal Q}_{IJKL}^{EF}(p;q,r') \label{s1}
\ee
as a {\it necessary} condition.
But it is easily seen from (\ref{zzz}) that because of the 
$r$-dependence on ${\cal Q}$ comes enterely through the $\delta_r$ factor,
the $p$-dependence on $\omega^{(p)}$ must be in the same way and
taking into account (\ref{simw}) the general form of it must be
\be
\omega^{(p)}_{IJK}{}^L = z_{IJK}{}^L + \delta_p \; z_{IJK}{}^L
\label{s2}
\ee 
with $z_{IJK}{}^L$ arbitrary, possibly dependent on $n$.
Then plugging (\ref{s2}) in (\ref{j32}) and using the identity
\be
\sum_{p=0}^n \; {2n+1\choose p}\; (-)^{ \frac{p\tilde p}{2} } = 
\eta\; 2^n \label{s3}
\ee
the following condition is obtained
\be
z_{IJK}{}^L + \eta\; z_{IKJ}{}^L = 0 \label{s4}
\ee
which in turn as before fixes the form of the $z$'s to be
\be
z_{IJK}{}^L = m_{IJK}{}^L - \eta \; m_{IKJ}{}^L
\label{s5}
\ee
with $m_{IJK}{}^L$ constants coefficients.

Finally equation (\ref{j12}) gives
\be
m_{IJK}{}^L = \delta_I^L\; \mu'_{JK}
\ee
with $\mu'$ an arbitrary $N_D\times N_D$ matrix. 
So the coefficients $\omega$'s can be recast in the final form
\be
\omega^{(p)}_{IJK}{}^L = \delta_I^L\; \mu_{JK} + \delta_p \; \delta_J^L\;
\mu_{IK}
\ee
with the matrix $\mu$ satisfying the symmetry property
\be
\mu_{IJ} = -\eta\; \mu_{JI} \label{simu}
\ee
It is straightforward to prove that they are solution of (\ref{j22}) for
arbitrary $\mu$.
This arbitrariness of the matrix $\mu$ should be waited from the fact 
that the algebra as usual is defined up to changes of basis; in fact under 
a change of basis
\bea
Q_I &\longrightarrow& P^J{}_I\; Q_J\nn\\
Z^{(p)}_{IJ}&\longrightarrow& P^K{}_I \; P^L{}_J\; Z^{(p)}_{KL}
\eea
the algebra remains invariant of form it is made the replacement
\be
\mu \longrightarrow P^{\rm t}\;\mu \; P \label{mu}
\ee
Then depending on the value of $\eta$ (and so on $n$) the matrix
$\mu$ can be taken in some standard form allowed by the
transformation (\ref{mu}) according to (\ref{simu}).

The final form of the $Z$-algebra is
\bea
\left[ Q_K^\Lambda , Z_{IJ(\mu_p )}^{(p)}\right] &=&
({\Gamma_{(\mu_p )}})^\Lambda{}_\Omega\; \left(
\mu_{JK}\; Q_I^\Omega + \delta_p\; \mu_{IK}\; Q_J^\Omega \right) \\
\left[ Z_{IJ(\mu_p )}^{(p)}, Z_{KL(\nu_q)}^{(q)}\right] &=& 
\left[ \left(\mu_{LI}\;\sum_{r=0}^n\;\frac{1}{r!}\;
\cC^{(\rho_r)}_{(\mu_p)(\nu_q)} Z_{KJ(\rho_r)}^{(r)}
\right) + \delta_p \; \left( I\leftrightarrow J\right)\right]\nn\\
&+&\delta_q\; \left[ K\leftrightarrow L\right] 
\eea

Let us explicitly write the superalgebras obtained in some particular cases.
\bigskip

\noindent{\underline{$n=5, \; N = 1$}}

As said in the introduction this eleven dimensional algebra could be of
relevance in M-theory; the $Z^{(p)}$- operators are presumibly related (as it 
is showed for particular cases where they effectively behave as {\it central} 
charges) to different charges associated to states that classically are 
p-branes like solutions of its low energy effective theory, eleven
dimensional supergravity. 
Taking into account the symmetry constraint (\ref{simZ}) there exist
charges for $p=1,2,5$ (commonly associated with the massless superparticle, 
supermembrane and superfivebrane of M-theory); the superalgebra is 
\bea
\left\{ Q^\Lambda , Q^\Omega \right\} &=& 
(\Gamma^{M} \cC^{-1} )^{\Lambda\Omega}\; Z_M^{(1)} - 
{1\over 2} \; ( \Gamma^{MN} \cC^{-1} )^{\Lambda\Omega}\; Z_{MN}^{(2)}\nn\\ 
&+& {1\over 5!}\; (\Gamma^{(\mu_5 )} \cC^{-1} )^{\Lambda\Omega}\;
Z_{(\mu_5)}^{(5)} \label{firstd=11}\\
\left[ Q^\Lambda , Z_{(\mu_p )}^{(p)}\right] &=&
2\;\mu\; ({\Gamma_{(\mu_p )}})^\Lambda{}_\Omega\;Q^\Omega\\
\left[ Z_{M}^{(1)}, Z_{N}^{(1)}\right] &=& 4\;\mu\; Z_{MN}^{(2)}\\
\left[ Z_M^{(1)} , Z^{(2)}_{N_1 N_2} \right] &=& 4\;\mu \left( 
\eta_{M N_1}\;  Z_{N_2}^{(1)} - \eta_{M N_2} \; Z_{N_1}^{(1)}\right)\\
\left[ Z_M^{(1)} , Z^{(5)}_{(\nu_5)} \right] &=& 4\;\mu\; {1\over 5!}\;
\epsilon^{(\rho_5}{}_{M\nu_5)}\; Z_{(\rho_5 )}^{(5)}\\
\left[ Z_{MN}^{(2)}, Z_{PQ}^{(2)} \right] &=& 4\;\mu\;\left(
\eta_{MQ}\; Z_{NP}^{(2)} + \eta_{NP}\; Z_{MQ}^{(2)} \right) - 
(P\leftrightarrow Q)\\
\left[ Z_{M_1 M_2}^{(2)} , Z^{(5)}_{(\nu_5)} \right] &=& 4\;\mu \;\left[
\left( \eta_{M_2 N_1}\; Z_{M_1 N_2\dots N_5}^{(5)} + 
{\rm cyclic}(N_1\dots N_5)\right) - \left( M_1 \leftrightarrow M_2
\right)\right]\\
\left[ Z_{(\mu_5)}^{(5)} , Z^{(5)}_{(\nu_5)} \right] &=& 4\;\mu  \biggl(
- \epsilon^{(R_1}{}_{\mu_5\nu_5)}\; Z_{R_1}^{(1)} + 
\sum_{\sigma ,\tau \in {\cal P}_5 }\; (-)^{\sigma+ \tau}\; 
\biggl({1 \over 24}\;
\prod_{l=1}^4 \; \eta_{M_{\sigma(l)} N_{\tau(l)}} 
Z_{M_{\sigma(5)} N_{\tau(5)}}^{(2)} \nn\\
&+& {1\over {72\; 5!}} 
\prod_{l=1}^2  \eta_{M_{\sigma(l)} N_{\tau(l)}}
\epsilon^{(\rho_5}{}_{M_{\sigma(3)}\dots
M_{\sigma(5)}N_{\tau(3)}
\dots N_{\tau(5)})} Z_{(\rho_5 )}^{(5)} \biggr)\biggr)\label{lastd=11}
\eea
\bigskip

\noindent{\underline{$n=4, \; N = 2$}}
\footnote{
Because $\;\eta=+1\;$ if $\;n=4 \; \rm{mod}\; 4$, there exists no non trivial 
$N=1$ extension (see (\ref{simu})) and the $Z$'s on the RHS of (\ref{qq})
behave like ``central" charges, but of course that $\mu = 0$ is always the 
trivial choice for any $n, N$.
}

The matrix $\mu$ is written as $\;\mu_{IJ} = \mu\; \epsilon_{IJ}$ and 
the symmetry constraint (\ref{simZ}) does not rule out any $p$-charge; 
being $\;\delta_0 = \delta_1 = -\delta_2 = -\delta_3 = \delta_4 = 1\;$, 
I write $Z_{IJ(\mu_0 )}^{(0)} \equiv Z_{IJ}$ and 
$\; Z_{IJ(\mu_p )}^{(p)}\equiv \epsilon_{IJ}\; Z_{(\mu_p)}^{(p)}\;\;$ 
for $p=2,3\;$, being symmetric for $p=0, 1, 4$. 
With these conventions the algebra is
\footnote{
Let us notice in particular that the subalgebra for $p=0$ in (\ref{z0alg})
is isomorphic to $sp(2, \Re )$ (take $J_3\sim Z_{12} /2\mu\; ,\;
J_+\sim Z_{11} /2\mu \; , \; J_- \sim Z_{22} /2\mu\;$ ; then 
$[J_3 , J_{\pm}] = \pm\; J_\pm \; ,  [J_+ , J_- ] = -2\;J_3 \;$).   
Its appearence is natural from the fact that this is the group of
automorphisms of the algebra since their transformations according to 
(\ref{mu}) leave invariant the matrix $\mu$.
}
\bea
\left\{ Q^\Lambda_I , Q_{J;\Omega} \right\} &=& 
\delta^\Lambda_\Omega\; Z_{IJ} + (\Gamma^M )^\Lambda{}_\Omega\; Z_{IJ(M)}^{(1)} 
+{1\over 4!} \; ( \Gamma^{\mu_4} )^\Lambda{}_\Omega\;
Z_{IJ(\mu_4)}^{(4)}\nn\\
&-& \epsilon_{IJ}\;\left({1\over 2!}\; (\Gamma^{MN})^\Lambda{}_\Omega\; 
Z_{(MN)}^{(2)} +  {1\over 3!}\; (\Gamma^{(\mu_3 )})^\Lambda{}_\Omega\;
Z_{(\mu_3)}^{(3)}\right)\\
\left[ Q^\Lambda_K , Z_{IJ(\mu_p )}^{(p)}\right] &=&
\mu\; \left( {\Gamma_{(\mu_p )}})^\Lambda{}_\Omega\; 
(\epsilon_{JK}\; Q^\Omega_I + \epsilon_{IK}\; Q^\Omega_J \right)\\
\left[ Q^\Lambda_I , Z_{(\mu_p )}^{(p)}\right] &=&
- \mu\; ({\Gamma_{(\mu_p )}})^\Lambda{}_\Omega\;Q^\Omega_I \\
\left[ Z_{IJ} , Z_{KL(\mu_p)}^{(p)}\right] &=& \mu\;\left( \epsilon_{LI}
Z_{KJ(\mu_p )}^{(p)} + \epsilon_{LJ}\; Z_{KI(\mu_p )}^{(p)}\right) + 
(K\leftrightarrow L)\label{z0alg}\\ 
\left[ Z_{IJ} , Z_{(\mu_p)}^{(p)}\right] &=& 0\\
\left[ Z_{IJ(M)}^{(1)}, Z_{KL(N)}^{(1)}\right] &=& \mu\; \biggl( 
\eta_{MN}\; ( \epsilon_{LI}\; Z_{KJ} + \epsilon_{LJ}\; Z_{KI} + 
\epsilon_{KI}\; Z_{LJ} + \epsilon_{KJ}\; Z_{LI} )\nn\\
&+&\ 2\; (\epsilon_{IL}\; \epsilon_{JK} + \epsilon_{IK}\;
\epsilon_{JL} )\; Z_{(MN)}^{(2)}\biggr)\\
\left[ Z_{IJ(M)}^{(1)} , Z^{(2)}_{(N_1 N_2)} \right] &=& 
2\;\mu \left( \eta_{M N_2} \;Z_{IJ(N_1 )}^{(1)}
- \eta_{M N_1}\;  Z_{IJ(N_2 )}^{(1)} \right)\\
\left[ Z_{IJ(M)}^{(1)} , Z^{(3)}_{(\nu_3)} \right] &=& -2\;\mu\; 
Z_{IJ(M\nu_3 )}^{(4)}\\
\left[ Z_{IJ(M)}^{(1)} , Z^{(4)}_{KL(\nu_4)} \right] &=& \mu\; \biggl( 
(\epsilon_{IL}\; \epsilon_{JK} + \epsilon_{IK}\; \epsilon_{JL} )\;
(\eta_{MN_1}\; Z^{(3)}_{N_2 N_3 N_4} - \eta_{MN_2}\; Z^{(3)}_{N_3 N_4
N_1}\nn\\
&+&\eta_{MN_3}\; Z^{(3)}_{N_4 N_1 N_2} 
- \eta_{MN_4}\; Z^{(3)}_{N_1 N_2 N_3})\nn\\
&+& {i\over 4!}\; \epsilon^{(\rho_4 }{}_{M\nu_4 )}\;
(\epsilon_{IL}\; Z_{KJ(\rho_4)}^{(4)} + \epsilon_{JL}\;
Z_{KI(\rho_4)}^{(4)})\biggr) + \left( K\leftrightarrow L\right)\\ 
\left[ Z_{(MN)}^{(2)}, Z_{(PQ)}^{(2)} \right] &=& 2\;\mu\;\left(
\eta_{MP}\; Z_{(NQ)}^{(2)} + \eta_{NQ}\; Z_{(MP)}^{(2)}\right) -
(P\leftrightarrow Q)\\
\left[ Z_{(M_1 M_2 )}^{(2)} , Z^{(3)}_{(\nu_3)} \right] &=& 2\;\mu\;
\left( \eta_{M_1 N_1}\; Z_{(M_2 N_2 N_3 )}^{(3)} + 
{\rm cyclic}(N_1 N_2 N_3)\right) - \left( M_1 \leftrightarrow M_2\right)\\
\left[ Z_{(\mu_2 )}^{(2)} , Z^{(4)}_{IJ(\nu_4)} \right] &=& 2\;\mu\;
\biggl( \eta_{M_1 N_1}\; Z_{IJ(M_2 N_2 N_3 N_4)}^{(4)} 
- \eta_{M_1 N_2}\; Z_{IJ(M_2 N_3 N_4 N_1)}^{(4)}\nn\\ 
&+& \eta_{M_1 N_3}\; Z_{IJ(M_2 N_4 N_1 N_2)}^{(4)} 
- \eta_{M_1 N_4}\; Z_{IJ(M_2 N_1 N_2 N_3)}^{(4)}\biggr)\nn\\ 
&-& \left( M_1 \leftrightarrow M_2\right)\\
\left[ Z_{(\mu_3)}^{(3)} , Z^{(3)}_{(\nu_3)} \right] &=& \mu \; \biggl(
\sum_{\sigma ,\tau \in {\cal P}_3}\; (-)^{\sigma +\tau}\;
\eta_{M_{\sigma (1)} N_{\tau (1)}}\; \eta_{M_{\sigma (2)} N_{\tau (2)}}\;
Z_{( M_{\sigma (1)} N_{\tau (1)} )}^{(2)}\nn\\
&-& {i\over 3}\; \epsilon^{(\rho_3}{}_{\mu_3\nu_3 )}\;
Z^{(3)}_{(\rho_3)}\biggr) \\
\left[ Z_{(\mu_3)}^{(3)} , Z^{(4)}_{IJ(\nu_4)} \right] &=& 2\; \mu \;
\biggl( \sum_{\tau \in {\cal P}_4 }\; (-)^\tau \; \prod_{l=1}^3 \;
\eta_{M_l N_{\tau(l)}}\; Z_{IJ(N_{\tau
(4)})}^{(1)} \nn\\
&+& \frac{i}{288} \sum_{\sigma\in{\cal P}_3 \atop \tau\in{\cal
P}_4} (-)^{\sigma +\tau}  \eta_{M_{\sigma(1)} N_{\tau(1)}} 
\epsilon^{(\rho_4}{}_{M_{\sigma(2)}M_{\sigma(3)} N_{\tau(2)} N_{\tau(3)}
N_{\tau(4)})} Z_{IJ(\rho_4 )}^{(4)} \biggr)\nn\\
& & \\
\left[ Z_{IJ(\mu_4)}^{(4)} , Z^{(4)}_{KL(\nu_4)} \right] &=& \mu \;
\biggl\{ \epsilon_{LI}\;\biggl( \;\sum_{\tau \in {\cal P}_4 }\; (-)^\tau
\; \prod_{l=1}^4 \;\eta_{M_l N_{\tau(l)}}\; Z_{KJ} \nn\\
&+& \frac{1}{8}\; \sum_{\sigma ,\tau\in{\cal P}_4}\; (-)^{\sigma
+\tau} \;\prod_{l=1}^2 \;\eta_{M_{\sigma(l)} N_{\tau(l)}}\;  
Z^{(4)}_{KJ ( M_{\sigma(3)}M_{\sigma(4)} N_{\tau(3)}
N_{\tau(4)})}\nn\\
&+& \frac{\epsilon_{KJ}}{6}\; \sum_{\sigma ,\tau\in{\cal P}_4}\; 
(-)^{\sigma +\tau} \;\biggl( \prod_{l=1}^3 \;
\eta_{M_{\sigma(l)} N_{\tau(l)}}\;
Z^{(2)}_{(M_{\sigma(4)}N_{\tau(4)})}\nn\\
&-& \frac{i}{36}\; \eta_{M_{\sigma(1)} N_{\tau(1)}}\; 
\epsilon^{(\rho_3}{}_{ M_{\sigma(2)}M_{\sigma(3)}M_{\sigma(4)} 
N_{\tau(2)} N_{\tau(3)}N_{\tau(4)})}\; Z^{(3)}_{(\rho_3
)} \biggr)\nn\\
&-& i\; \epsilon^{(R}{}_{\mu_4\nu_4 )}\;
Z^{(1)}_{KJ(R)} \biggr) + (I\leftrightarrow J ) \biggr\} + 
\left( K\leftrightarrow L\right)
\eea

\noindent{\underline{$n=1,\; N\; \rm{arbitrary} $} }

This three dimensional algebra presents one scalar charge 
$\; Z_{IJ(\mu_0 )}^{(0)} \equiv Z_{IJ} = - Z_{JI}\;$ and a vector one 
$\; Z_{IJ(M)}^{(1)} =  Z_{JI(M)}^{(1)}$. 
The matrix $\mu$ is symmetric and by definiteness I take it to be 
pseudorthogonal with signature $(N-d,d)$, $\mu_{IJ} = \eta_{IJ}$.  
The algebra reads
\bea
\left\{ Q^\Lambda_I , Q_{J;\Omega} \right\} &=& \delta^\Lambda_\Omega\; 
Z_{IJ} + (\Gamma^M )^\Lambda{}_\Omega\; Z_{IJ(M)}^{(1)}\\
\left[ Q^\Lambda_K , Z_{IJ}\right] &=&
\eta_{JK}\; Q^\Lambda_I - \eta_{IK}\; Q^\Lambda_J\\
\left[ Q^\Lambda_K , Z_{IJ(M)}^{(1)}\right] &=&
({\Gamma_M})^\Lambda{}_\Omega\; 
\left( \eta_{JK}\; Q^\Omega_I + \eta_{IK}\; Q^\Omega_J \right)\\
\left[ Z_{IJ} , Z_{KL}\right] &=& \eta_{LI}\; Z_{KJ} -
\eta_{LJ}\; Z_{KI} - (K\leftrightarrow L )\\
\left[ Z_{IJ} , Z_{KL(M)}^{(1)}\right] &=& \eta_{LI}\; Z_{KJ(M)}^{(1)}
- \eta_{LJ}\; Z_{KI(M)}^{(1)} + (K\leftrightarrow L )\\
\left[ Z_{IJ(M)}^{(1)} , Z^{(1)}_{KL(N)} \right] &=& 
\left( \eta_{LI}\; \left( \eta_{MN}\; Z_{KJ} - \epsilon^{(R}{}_{MN)}
\;  Z_{KJ(R)}^{(1)} \right) + (I\leftrightarrow J) \right)\nn\\
&+&(K\leftrightarrow L)
\eea

\section*{Conclusions}

I have presented in this paper an $N$-extended superalgebra by $p$-form  
tensor operators for any dimension in which Majorana spinors exists, in
some sense a generalization of early work in reference \cite{holpro} where
the minimal case $N=1$ was considered. 
\footnote{
Also the case of dimensions $D=9 \; \rm{mod}\; 8$ was not considered
there.
} 
This is only a first step however, in the direction of a generalization
of this work that should contain spin operators other than the
supersymmetric charges. 
This inclusion is motivated mainly from the fact that formulations of a
$N$-extended target superspace that add new degrees of freedom (the
coordinates corresponding to the new generators) could be of most
importance in the construction of super $p$-branes actions \cite{sez2}.
The fact that the algebra given in (\ref{firstd=11})-(\ref{lastd=11}) does 
not reduce to the $D=11$ algebra discovered in \cite{sez1}) seems to give
a hint that this generalization should exist and be parameter-dependent. 
Also it should be interesting to work out the representations as well as
field-theoretic realizations of these algebras.
Finally it is worth to remark that even dimensional results can be
straightforwardly got from naive dimensional reduction from the results
presented here; however not all the possibilities might be contemplated
this way because more covariance constraints that the really needed ones
are included following this route, in particular in writting equations
like (\ref{qz}). 
I hope to address some of these questions in a near future. 
\bigskip

I thank to Jos\'e Edelstein for bringing to my attention references 
\cite{sez1}, \cite{sez2} (from which I learnt about the existence of
\cite{holpro}) and for useful comments.

\appendix

\section{Gamma matrices and all that}
\label{apa}

I think is worth to say that much of the material here can be found
(among many others certainly) in \cite{susy}. 
As it is well known a set of $2n + 1$ matrices 
$\{\Gamma^M = ((\Gamma^M)^\Lambda{}_\Omega )\; ; M= 0,1, \dots ,2n, \; 
\Lambda , \Omega= 1, \dots , 2^n \}$ of dimension $2^n \times 2^n$ that 
satisfy the Clifford algebra
\be
\left\{ \Gamma^M ,\Gamma^N \right\} = 2\;\eta^{MN}\; 1 \;\;,\;\; 
\eta \equiv {\rm diag}(-1, 1,\dots, 1) 
\label{gama1}
\ee
with 
\be
\Gamma^{2n} =  i^{n+1}\; \Gamma^0 \;\Gamma^1 \dots \Gamma^{2n-1} = 
\left( \begin{array}{cc}  1 & 0 \\
           0 & -1  \end{array}\right)  \label{gama2}
\ee
and the further properties
\begin{eqnarray}
{\Gamma^M}^t &=& (-)^M \;\Gamma^M \nn\\
{\Gamma^M}^* &=& (-)^M \;\eta^{MM}\; \Gamma^M  \label{gama3}
\end{eqnarray}
can be constructed by induccion for $n=k$ if they exist for $n=k-1$. 
In fact if $\{ \gamma^\mu , \;\mu =0,\dots , 2(k-1) \}$ is a set of
$2(k-1)+1=2k-1$ matrices of dimension $2^{(k-1)} \times 2^{(k-1)}$
satisfying relations (\ref{gama1}), (\ref{gama2}), (\ref{gama3}), then 
the $2^k \times 2^k$ dimensional matrices defined by 
\bea
\Gamma^\mu &=& \left( \begin{array}{cc} 0 &  \gamma^\mu \\
                \gamma^\mu & 0  \end{array}\right) 
\;\; ,\;\;  \mu=0, 1 , \dots, 2k - 2 \nn\\
\Gamma^{2k-1} &=& i \;  \left( \begin{array}{cc} 0 &  1\\
                                 -1 & 0  \end{array}\right) \nn\\
\Gamma^{2k} &=&  i^{k+1} \;\Gamma^0 \Gamma^1 \dots \Gamma^{2k-1}
\label{gama}
\eea
also satisfy them as it is straightforwardly showed.

An initial set for $n=1$ can be taken as 
\be
\sigma^0 = i\;\sigma_1 \;\; , \;\; \sigma^1 = \sigma_2 \;\; ,\;\;  
\sigma^2 = \sigma_3 \label{n=1}
\ee
where $\sigma_i$ are the standard Pauli matrices.
However it can be useful in some cases to start with $n=2$ and the
standard four dimensional matrices
\begin{eqnarray}
\gamma^{\mu}_s &=& i \; \left( \begin{array}{cc}  0 &  \sigma^\mu \\
           - \overline{\sigma}^\mu & 0 \end{array}\right) \;\; ,\;\;
\mu = 0, 1, 2, 3\nn\\
\gamma^{4}_s &=& \gamma_5 =  \left( \begin{array}{cc} 1 &  0\\
                                     0 & -1 \end{array}\right) 
\end{eqnarray}
where $( \sigma^\mu ) = ( 1 , \sigma_1 , \sigma_2 , \sigma_3 )$ and  
$( {\overline\sigma}^\mu ) = ( - 1 , \sigma_1 , \sigma_2 , \sigma_3 )$. 
These matrices and the ones obtained from the construction given in
(\ref{gama}) starting from (\ref{n=1}) are of course related by a
similarity transformation that for completeness I quote
\bea
\gamma^\mu &=& S\;\gamma^\mu_s\; S^{-1} \nn\\
S &=& \frac{1}{\sqrt 2} \left( \begin{array}{cccc}  1&-1&0&0\\
                                 1& 1&0&0\\
                                  0&0&1&1\\
                                  0&0&1&-1 \end{array}\right) = (S^t)^{-1}
\eea

I oftenly use in the paper the shorthand notation 
\be
(\mu_p) \equiv (M_1 \dots M_p )
\ee
for a set of completely antisymmetric indices $\{ M_i = 0,1,\dots ,2n\}$;
in particular 
\bea
\epsilon^{M_1 \dots M_{2n+1}} &\equiv& \epsilon^{ (\mu_{2n+1} )} \;\;\; ,
\;\;\; \epsilon^{01\dots 2n} \equiv +1 \nn\\
\epsilon^{(\mu_p }{}_{\Sigma_{\tilde p})}\;
\epsilon^{(\Sigma_{\tilde p}}{}_{\nu_p)} &=& p!\; {\tilde p}!\; 
\delta^{(\mu_p}{}_{\nu_p)} 
\eea
where $ {\tilde p} \equiv  2n + 1 - p $ and the $\delta$ that appears is
the identity matrix in the space of antisymmetric tensors
\be
\delta^{(\mu_p )}{}_{(\nu_p)}\; A^{(\nu_p)} \equiv A^{(\mu_p)}
\label{deltadef}
\ee
As usual the raising and lowering of indices is covariantly
made by means of the $\eta_{MN}$ - matrix.
 
The sets of completely antisymmetrized products
\footnote{
By convention, $\Gamma^{(\mu_0 )} \equiv 1 $
}
\be
\Gamma^{(\mu_p )} \equiv 
\Gamma^{M_1\dots M_p} = \frac{1}{p!}
\sum_{\sigma\in{\cal P}_p}\; (-)^\sigma\;\Gamma^{M_{\sigma(1)}}
\dots\Gamma^{M_{\sigma(p)}}\;\;\; ,\;\;\; p=0, 1,\dots, 2n+1 
\ee
where ${\cal P}_p$ is the group of permutations of $p$ elements and 
$(-)^\sigma$ the sign of the corresponding element $\sigma$, are complete
in the $2^{2n}$ dimensional vector space of $2^n \times 2^n $ matrices.
However they are not independent as can be seen from the total number of
matrices, $2^{2n+1}$; in fact the sets with $n+1 \le p \le 2n+1$ and 
$0\le p \le n $ are duality-related
\footnote{
I will consistently ommit the $n$-dependence in all factors and phases.
}
\bea
\Gamma^{(\mu_p )} &=& \alpha_p \; \epsilon^{(\mu_p}{}_{\nu_{\tilde p})}\; 
\Gamma^{(\nu_{\tilde p})} \nn\\
\alpha_p &=& (-i)^{n+1} \; 
\frac{ (-)^{ \frac{\tilde p}{2} ({\tilde p}-1)} }{{\tilde p}!}\label{dual}
\eea
Then is enough for a basis to take $p=0, 1, \dots , n \; $, giving a total
of $2^{2n}$ matrices as it should.
In particular for $p=2$ the matrices $ S_{MN} = \frac{1}{2} \Gamma_{MN}$
are the generators of ${\rm Spin}(1,2n)$ in the spinor representation
obeying the standard algebra (\ref{loralg}).
A useful relation I quote is
\be
{1\over p!}\;\Gamma^{(\mu_p )}\; \Gamma_{(\mu_p )} =  
(-)^{ {p\over 2}(p-1)}\;{2n+1\choose p}\; 1
\ee

The charge conjugation matrix $\cC$ is defined by  
\be
\cC \equiv i^n \; \Gamma^1 \Gamma^3 \Gamma^5 \dots \Gamma^{2n-1} 
= ( \cC_{\Lambda\Delta}) = ( \cC^{\Lambda\Delta})
\ee
In a basis like (\ref{gama}) it verifies the relations
\bea
\cC &=& \eta\;\cC^{-1} = \eta \;\cC^t = \cC^* = \cC^\dagger\nn\\
\eta &=& (-)^{\frac{n}{2} (n+1)}
\eea
The further properties are also important
\bea
(\Gamma^{(\mu_p )}\;\cC^{-1})^t &=&
\delta_p\;\Gamma^{(\mu_p)}\;\cC^{-1}\nn\\ 
\cC \; \Gamma^{(\mu_p )} \cC^{-1} &=& \eta\; \delta_p
\;{\Gamma^{(\mu_p)}}^t\nn\\
\delta_p &=& \eta\; (-)^{\frac{p\tilde p}{2}} = \delta_{\tilde p}
\label{gamaprop1}
\end{eqnarray}
In particular for $p=2$  and  any $n$
\be
\cC \; S_{MN} \;\cC^{-1} = - (S_{MN})^t  \;\longleftrightarrow\; 
\cC \; S(\omega ) \;\cC^{-1} = (S(\omega)^t )^{-1}   
\ee
where $\; S(\omega)\equiv e^{\frac{1}{2}\omega^{MN} S_{MN}}\;$ is an
arbitrary element of ${\rm Spin}(1,2n)$ in the spinor representation
parametrized by the coefficients $\{ \omega_{MN}= - \omega_{NM} \}$,
being manifest the (defining) fact that $\cC$ is the intertwining matrix
between the spinor representation and its transpose inverse one, 
fact that allows to raise and lower indices with $\cC$ in a covariant way.

Another identity of particular interest in the present work is 
\bea
{1\over p!}\;(\Gamma^{(\mu_p )}\;\cC^{-1})^{\Lambda\Omega}
\; (\Gamma_{(\mu_p)}\;\cC^{-1})^{\Delta\Gamma} &=& 
(-)^{{p\over 2}(p-1)}\; {2n+1\choose p} \; f^{-1}\; 
\biggl( \delta_{p,0}\; 2^n\; \cC^{\Lambda\Omega ; \Delta\Gamma}\nn\\
&+& \cC^{\Delta\Omega ; \Lambda\Gamma} + \eta\; \delta_p\; 
\cC^{\Lambda\Delta ; \Omega\Gamma}\biggr)\\
f &=& 2^{2n} + \eta\; 2^n - 2
\label{gamaprop2}
\eea 
valid for $n>1$, where the tensor 
\be
\cC^{\Lambda\Omega ; \Delta\Gamma} \equiv 
(2^n + \eta )\;  \cC^{\Lambda\Omega}\; \cC^{\Delta\Gamma} - 
\cC^{\Lambda\Delta}\; \cC^{\Omega\Gamma}-
\cC^{\Delta\Omega}\; \cC^{\Lambda\Gamma}
\ee
has the properties
\bea
\cC^{\Lambda\Omega ; \Delta\Gamma} &=& \cC^{\Delta\Gamma ;\Lambda\Omega} =
\eta\; \cC^{\Lambda\Omega ; \Gamma\Delta} =
\eta \;\cC^{\Omega\Lambda ; \Delta\Gamma}\nn\\
\cC^{\Lambda\Omega ; \Delta\Gamma}\; \cC_{\Omega\Delta} &=& 0 \nn\\
\cC^{\Lambda\Omega ; \Delta\Gamma}\; 
\cC_{\Lambda\Omega} &=& f\; \cC^{\Delta\Gamma}
\eea

The other basic representation of the ${\rm Spin}(1, 2n)$ generators
satisfying (\ref{loralg}) is the $2n+1$ -dimensional vector representation
defined by
\be
(V_{MN})^P{}_Q = \delta^P_M\; \eta_{NQ} - \delta^P_N\; \eta_{MQ}
\ee
Under ${\rm Spin}(1, 2n)$ the $\Gamma^{(\mu_p)}$'s transform like 
\be
{S(\omega)}^{-1}\; \Gamma^{(\mu_p)}\; S(\omega) = 
V(\omega)^{(\mu_p)}{}_{(\nu_p)}\; \Gamma^{(\nu_p)} \label{gamalor}
\ee
where 
\be
V(\omega )^{(\mu_p)}{}_{(\nu_p)} = {1\over p!}\; 
\sum_{\sigma\in{\cal P}_p}\; (-)^\sigma\;V(\omega 
)^{M_{\sigma(1)}}{}_{N_1}
\dots V(\omega )^{M_{\sigma(p)}}{}_{N_p}\;\;\; ,\;\;\; p=0, 1,\dots, 2n+1 
\label{vdef}
\ee
is the order $p$ completely antisymmetric tensor representation of 
${\rm Spin}(1, 2n)$, being $V(\omega)\equiv e^{\frac{1}{2}\omega^{MN}
V_{MN}}$ a ${\rm Spin}(1, 2n)$ element in the vector representation.

\section{Invariant tensors and gamma products}
 
I introduce here Clebsh-Gordan-like invariant tensors that decompose the
product of completely antisymmetric representations of $SO(1, 2n)$, 
normalized in a convenient way.
The tensor $C^{(\rho_r)}_{(\mu_p)(\nu_q)}$ is defined to be zero if
$p,q,r$ lie outside the completely symmetric region 
\bea
0&\le& p,q,r \le 2n+1 \;\; , \;\; p+q+r \;\; {\rm even}\nn\\ 
p+q&\geq& r \;\; ,\;\; p+r\geq q \;\; ,\;\;  r+q\geq p  
\label{r}
\eea
where it has the form
\begin{eqnarray}
C^{(\rho_r)}_{(\mu_p)(\nu_q)} &\equiv& 
\gamma_{pq}^r\; 
\sum_{
{\sigma\in {\cal P}_p \atop 
  \tau\in {\cal P}_q }\atop
\delta\in {\cal P}_r
} 
\; (-)^{\sigma+\tau+\delta}
\prod_{l=1}^{\frac{p+q-r}{2}} \eta_{ M_{\sigma (l)} N_{\tau (l)} }
\prod_{k=1}^{\frac{p+r-q}{2}}
\delta^{ R_{\delta(k)} }_{ M_{ \sigma(\frac{p+q-r}{2}+k) } } 
\prod_{m=1}^{ \frac{q+r-p}{2} } \delta^{ R_{\delta(\frac{p+r-q}{2}+m)} }_{ 
N_{\tau(\frac{p+q-r}{2}+m)} }\nn\\
\gamma_{pq}^r &\equiv&
\frac{ (-)^{\frac{p+q-r}{4}( \frac{p+q-r}{2} + 1 + 2p )} }{
(\frac{p+r-q}{2})!\; (\frac{p+q-r}{2})! \; (\frac{q+r-p}{2})!}
\end{eqnarray}

Its key property is 
\be
\Gamma_{(\mu_p)}\; \Gamma_{(\nu_q)} =
\sum_{r=0}^{2n+1} \; \frac{1}{r!}\; C_{(\mu_p)(\nu_q)}^{(\rho_r)} \;
\Gamma_{(\rho_r)}\label{2gama}
\ee
whose proof by induction in both indices is left to the reader.
These tensors satisfy the following symmetry properties
\begin{eqnarray} 
C^{(\rho_r)}_{(\nu_q)(\mu_p)} &=& 
\sigma_{pqr}\; C^{(\rho_r)}_{(\mu_p)(\nu_q)}\nn\\
\sigma_{pqr}&\equiv& \eta\;\delta_p\;\delta_q\;\delta_r\label{simC}
\end{eqnarray}
I also introduce the ``dual" tensor defined as 
\be
{\tilde C}^{(\rho_r)}_{(\mu_p)(\nu_q)} \equiv
\frac{r!}{{\tilde r}!}\; \alpha_{\tilde r}
\;\epsilon^{ (\rho_r}{}_{ \Sigma_{\tilde r}) }\;
C^{ (\Sigma_{\tilde r}) }_{ (\mu_p) (\nu_q) } 
\ee
if $\; p, q , {\tilde r}\;$ 
belong to region (\ref{r}) and zero otherwise. 
They also satisfy {\it the same} symmetry property (\ref{simC})
\be
{\tilde C}^{(\rho_r)}_{(\nu_q)(\mu_p)} =
\sigma_{pqr}\;{\tilde C}^{(\rho_r)}_{(\mu_p)(\nu_q)}
\label{simCtilde}
\ee
By using the duality relation (\ref{dual}) and the definitions
above formula (\ref{2gama}) can be rewritten as
\be
\Gamma_{(\mu_p)}\; \Gamma_{(\nu_q)} =
\sum_{r=0}^{n} \; \frac{1}{r!}\; \cC_{(\mu_p)(\nu_q)}^{(\rho_r)} \;
\Gamma_{(\rho_r)}\label{2gamabis}
\ee
where I have introduced the tensor
\be
\cC_{(\mu_p)(\nu_q)}^{(\rho_r)} = \left\{ 
C_{(\mu_p)(\nu_q)}^{(\rho_r)} \;\;\; {\rm if}\;\;\; p+q+r \;\;\;{\rm even}
\atop {\tilde C}_{(\mu_p)(\nu_q)}^{(\rho_r)} \;\;\; {\rm if}\;\;\;
p+q+r\;\;\;{\rm odd}\right.
\ee
In particular for any $0\leq p, q \leq 2n+1$
\be
\rm{tr} \left( \Gamma_{(\mu_p)}\; \Gamma_{(\nu_q)}\right) =
2^n\; \cC_{(\mu_p)(\nu_p)}^{(\rho_0)} \;\delta_{p,q}
\ee
Finally from (\ref{simC}) and (\ref{simCtilde}) it follows
\be
{\cC}^{(\rho_r)}_{(\nu_q)(\mu_p)} = 
\sigma_{pqr}\; \cC^{(\rho_r)}_{(\mu_p)(\nu_q)}
\label{simcC}
\ee

\end{document}